\documentclass[conference, 10pt]{IEEEtran}
\IEEEoverridecommandlockouts
\usepackage{cite}
\usepackage{amsmath,amssymb,amsfonts}
\usepackage{algorithmic}
\usepackage{graphicx}
\usepackage{textcomp}
\usepackage{xcolor}
\def\BibTeX{{\rm B\kern-.05em{\sc i\kern-.025em b}\kern-.08em
    T\kern-.1667em\lower.7ex\hbox{E}\kern-.125emX}}

\usepackage{xspace} 
\usepackage{lmodern}
\newcommand{\etal}{\textit{et al.}\xspace} 
\usepackage{booktabs}
\usepackage{url}

\begin{document}

\title{Context-aware Container Orchestration in Serverless Edge Computing}

\author{\IEEEauthorblockN{Peiyuan Guan}
\IEEEauthorblockA{\textit{Dept. of Informatics } \\
\textit{University of Oslo}\\
Oslo, Norway \\
peiyuang@ifi.uio.no}
\and
\IEEEauthorblockN{Chen Chen}
\IEEEauthorblockA{\textit{Dept. of Computer Science and Technology} \\
\textit{University of Cambridge}\\
Cambridge, UK \\
cc2181@cam.ac.uk}
\and
\IEEEauthorblockN{Ziru Chen}
\IEEEauthorblockA{\textit{Dept. of  Electrical and Computer Engineering} \\
\textit{ Illinois Institute of Technology}\\
Chicago, USA \\
zchen71@hawk.iit.edu}
\and
\IEEEauthorblockN{Lin X. Cai}
\IEEEauthorblockA{\textit{Dept. of  Electrical and Computer Engineering} \\
\textit{ Illinois Institute of Technology}\\
Chicago, USA \\
lincai@iit.edu}
\and
\IEEEauthorblockN{Xing Hao}
\IEEEauthorblockA{\textit{School of Information Science and Technology
} \\
\textit{ Northwest Univesity}\\
Xi'an, China \\
xhao@nwu.edu.cn}
\and
\IEEEauthorblockN{Amir Taherkordi}
\IEEEauthorblockA{\textit{Dept. of Informatics } \\
\textit{University of Oslo}\\
Oslo, Norway \\
Amirhost@ifi.uio.no}

}

\maketitle

\begin{abstract}
Adopting serverless computing to edge networks benefits end-users from the pay-as-you-use billing model and flexible scaling of applications. This paradigm extends the boundaries of edge computing and remarkably improves the quality of services.
However, due to the heterogeneous nature of computing and bandwidth resources in edge networks, it is challenging to dynamically allocate different resources while adapting to the burstiness and high concurrency in serverless workloads.
This article focuses on serverless function provisioning in edge networks to optimize end-to-end latency, where the challenge lies in jointly allocating wireless bandwidth and computing resources among heterogeneous computing nodes.
To address this challenge, We devised a context-aware learning framework that adaptively orchestrates a wide spectrum of resources and jointly considers them to avoid resource fragmentation.
Extensive simulation results justified that the proposed algorithm reduces over 95\% of converge time while the end-to-end delay is comparable to the state of the art.
\end{abstract}


\section{Introduction}

The progressive development of Internet-of-Things (IoTs) and mobile devices has produced massive Internet traffic to cloud data centers, dramatically increasing the network pressure in backbone networks.
To cope with this issue \cite{chen2021performance}, many IoT applications can be offloaded from remote clouds to edge servers, reducing the traffic to remote clouds and improving the end-to-end latency.
Such a near-data paradigm largely mitigates the transmission problem that inhibits the deployment of delay-sensitive applications, e.g., online gaming, health care and autonomous vehicles \cite{Katare2023, Chen2022, Chen2021}.

Recently, serverless computing\cite{Li2023}, also known as Function-as-a-Service (Faas), endows edge computing with new inspirations.
Serverless applications are encapsulated in functions that are triggered by user events.
It hands over the server management to service providers and enables developers to focus only on application development \cite{chen2023crossedge}.
Developers upload these functions to the serverless platform and use the API server to execute their computation.
Given the promising prospect of serverless computing, many research works believe that bringing serverless computing to the network edge will realize new flexibility, efficacy and scalability.

\begin{figure}[t]
    \includegraphics[width=0.45\textwidth]{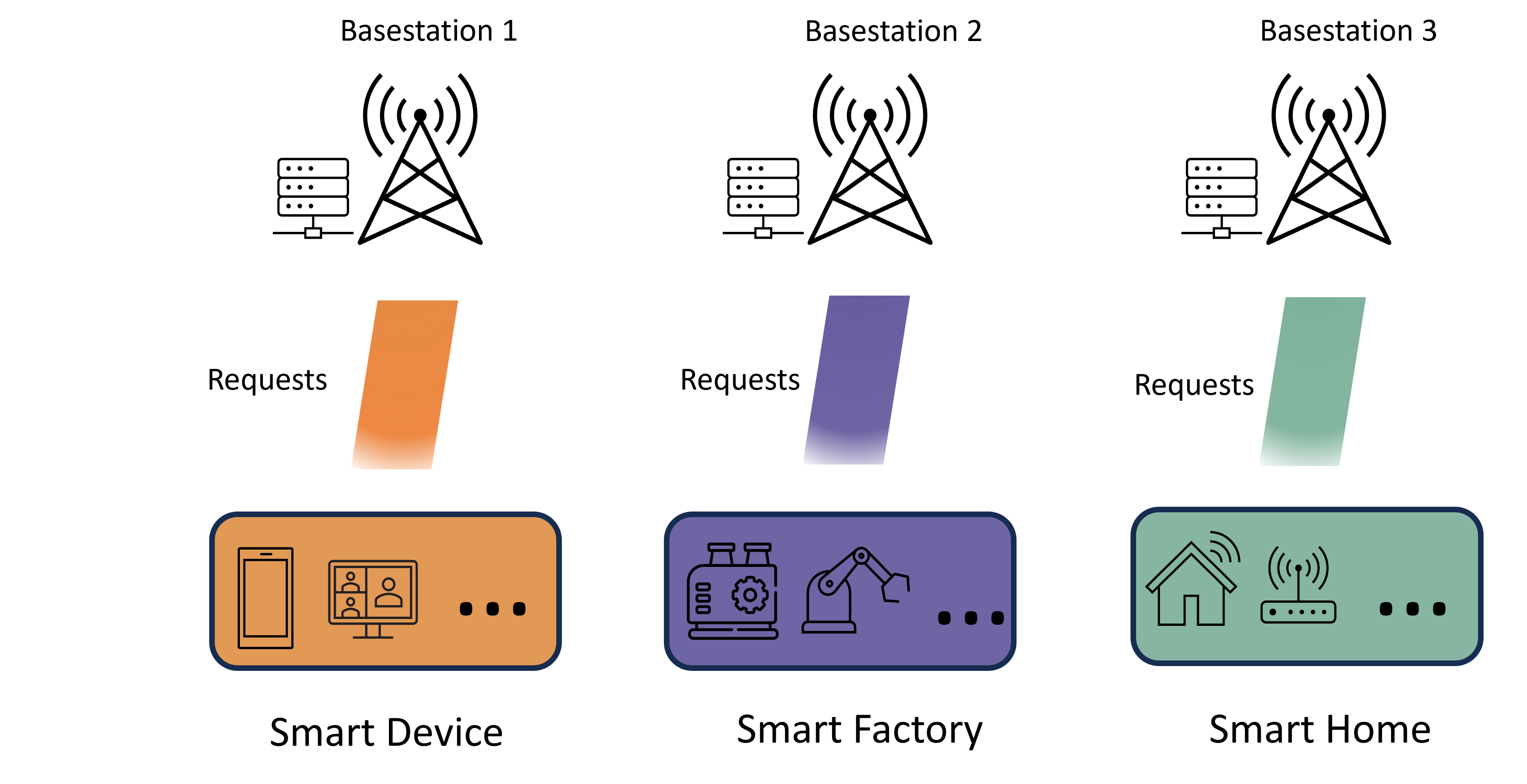}
    \centering
    \caption{Example of serverless edge computing}
    \label{fig:overview}
\end{figure}

A number of research works have been proposed to integrate the serverless paradigm into edge computing, e.g., \cite{Shang2023, Chen2023}.
However, existing solutions are not directly applicable because several inherent features of serverless edge computing have been overlooked.
First, edge servers are usually equipped with different amounts of communication and computation resources.
Such heterogeneity requires dynamic and adaptive resource allocation because the service latency can be bounded by a certain type of resource, e.g., communication bandwidth and processing capacity.
Second, the size and arrival pattern of serverless workloads can vary significantly due to the burstiness and concurrency over time, the resources must adapt to these changes in serverless workloads.

Since more and more base stations are increasingly equipped with computing capacity, as depicted in Figure~\ref{fig:overview}, service requests can be assigned to base station 2 if base station 1 is overloaded, aiming to achieve faster data transmission and processing.
In other words, how to jointly select the base stations and allocate bandwidth resources for low-transmission latency is non-trivial but complex.
Besides request assignment, extra efforts are still needed to allocate processing capacity for serverless functions on each edge server.
Since both computing resources are highly constrained at the network edge \cite{Xiao2024}, an efficient orchestration policy is vitally important for low-latency applications.

In this paper, we propose a serverless container placement and resource allocation framework for latency-sensitive applications, addressing the challenges mentioned above in serverless edge computing.
First, we formulate the problem as an Integer Programming problem with several constraints.
After that, we propose a context-aware neural network (CANN) based on the results of MIDACO\cite{MIDACOSO37:online}, which is a solver for numerical optimization problems, and compare the total delay and converge time with MIDACO and genetic algorithm.
Extensive numerical tests justified that MIDACO can obtain the global optimal fast and robustly on hundreds of benchmarks\cite{MartinSchlueter2012}.
Hence, we compare our algorithm with the results of MIDACO, justifying that our algorithm can achieve comparable performance.
Our main contributions are listed as follows.

\begin{enumerate}
\item We formulate a request distribution
problem as a Mixed Integer Linear Programming problem
that jointly considers transmission latency and processing latency with constraints of bandwidth and computing resources. We prove the
NP-hardness of the problem and propose an
online competitive algorithm for request distribution to solve the problem in polynomial time.

\item We propose an NN, trained by the results of MIDACO, that jointly considers the bandwidth and computing resources, dynamically adapting to the heterogeneity, burstiness, and high concurrency of serverless computing.

\item We conduct extensive experiments based on real-world datasets. 
To evaluate the performance, we compare the proposed algorithms with the best solution obtained by MIDACO, justifying that the proposed algorithm achieves a comparable performance for the total delay and the times of winning while the converge time decreases by about 95\%.
\end{enumerate}

The remainder of the paper is organized as follows. Section~\ref{section::relatedwork} gives an overview of the related work before presenting the system model and the problem in section~\ref{section::systemmodel} and \ref{section::problem}, respectively. 
After that, we present the proposed algorithm, the experiments and the conclusion.
\section{Related Work}\label{section::relatedwork}

A number of works investigate the container placement problem, optimizing various objectives such as latency, system cost and etc.

Shang \etal \cite{Shang2023} formulate a container placement and flow routing problem by considering the heterogeneity between edge nodes and the overhead of adopting serverless platform.
Extensive simulation results justify the efficacy of the proposed online algorithm by jointly considering delay, operating cost and data availability.
Hu \etal \cite{Hu2023} investigate the request scheduling problem in the context of Vehicle-Infrastructure Collaboration.
By using layer sharing and container sharing, this paper reduces the long-term system cost consisting of transmission cost, preparation cost, retention cost, computation cost and transfer cost.
Xiao \etal \cite{Xiao2024} study the cold-start problem in serverless edge computing, aiming to minimize the system cost incurred by cold-start, transmission and container caching.
Sahraei \etal \cite{Sahraei2023} propose an approach to improve the CPU utilization in Meta hyperscale private cloud. The proposed approach defers the invocations of delay-tolerant functions to off-peak hours and adopt a TCP-like congestion control policy to regulate the function execution.
Wang \etal \cite{Wang2024} adopt Markov decision process to model the service deployment problem in 6G networks, aiming to optimize the overall latency at a lower cost.
The paper uses a greedy algorithm to find service deployment in a multi-layer edge network where services are assigned to the nearest ancestor devices in the routing tree.

There are some other works investigating container placement in serverless computing.
ServerMore \cite{Suresh2021} proposes colocation of serverless applications with serverful VMs. 
By this means, ServerMore improves the resource utilization by up to $245\%$  with a minimal degradation of latency. 
Pan \etal \cite{Pan2022} resolve the container placement and retention problem by mapping it to the classic ski-rental problem. The proposed online algorithm opportunistically distributes request by jointly considering resource capacity and network latency.

However, none of the aforementioned work has considered the inherent nature of serverless workloads such as the various size of jobs and the resource contention in wireless transmission. By jointly considering those dynamics with the topology in edge computing, we set our work apart from existing approaches.

\section{System Model}\label{section::systemmodel}

We use $v \in \mathcal{V}$ to represent the set of edge nodes which is equipped with a processing capacity $C_v$.
Each edge node is a basestation that can process serverless requests.
Also, each edge node is equipped with a certain amount of wireless bandwidth which is denoted by $B_v$.
We use $k \in \mathcal{K}$ to denote a service request that requires a type $k$ container. For each service request $k$, we use $L_k$ to denote the size of the job. 
Also, we use a binary variable $x_{k,t}^v$ to represent the request $k$ is assigned to edge node $v$ in time slot $t$ when $x_{k,t}^v = 0$ and vice versa.
Let $c_{k,t}$ denote the required amount of hardware resources for request $k$ in time slot $t$. Also, we use $C_v$ to represent the total amount of hardware resources in edge node $v$.
Moreover, we use $b_{k,t}$ and $B_v$ to denote the deployment cost of running service request $k$ and the total budget allocated to edge node $v$, respectively.

In modeling the wireless communication scenario, the path-loss function as detailed in Equation~\ref{PL} and sourced from Giordani et al.~\cite{giordani2019path} is utilized to assess the impact of distance on each service request $k$.
\begin{equation} PL_i = 38.77 + 16.7*\log_{10}{d_k^v}+18.2*\log_{10}{f_k}
\label{PL} 
\end{equation} 
In Equation~\ref{PL}, $d_k^v$ is the distance (the units are
meters) between service request $k$ and the edge node $v$, whereas $f$ is the frequency
(we set this parameter as 5.9 GHz in this work) of the transmission signal. The units of path
loss computed by Equation~\ref{PL} are decibels (db). Path loss measurements obtained from Equation~\ref{PL} are expressed in decibels (dB). The coefficients 38.77, 16.7, and 18.2 are empirical values referenced from Giordani et al.~\cite{giordani2019path}.

Equation~\ref{acceptance signal power} determines the accepted signal power, measured in decibel-milliwatts (dBm), using $P_{k}$ to represent the signal power, which is $21$ dBm according to the wireless standard detailed in~\cite{giordani2019path}.
\begin{equation}
    P_{k}^{'} =P_k-PL_k.
    \label{acceptance signal power}
\end{equation}
It is important to recognize that both dB and dBm represent logarithmic scales; thus, $P_{k}^{'}$ is measured in dBm. To convert $P_{k}^{'}$ into milliwatts (mW), one should use Equation~\ref{signal power in mw}.
\begin{equation}
    P_{k}^{''} = 10^{ \frac{P_{k}^{'}}{10}}
    \label{signal power in mw}
\end{equation}
allows the computation of the signal-to-noise ratio (SNR) for each service request using Equation~\ref{SNR_i}
\begin{equation}
    SNR_k^v = \frac{P_{k}^{''}}{N_0*b_k^v}
    \label{SNR_i}
\end{equation}
where $N_{0}$ is the power of environment noise, and it equals $
10^{-11.4}$ mw~\cite{li2016uplink}.
The Shannon equation (see Equation~\ref{transmission speed}) calculates the transmission speed $TS_k^v$ for service request $k$ to edge node $v$, using the Signal-Noise Ratio ($SNR_k^v$) and Bandwidth ($b_k^v$, specified in MHz).
\begin{equation}
    TS_k^v = b_k^v*\log_{2}{(1+SNR_k^v)}
\label{transmission speed}
\end{equation}
To obtain the transmission time for request $k$, divide the data size $L_k$ for each service by the transmission speed $TS_k^v$.
\begin{equation}
    TT_k^v = \frac{L_k}{TS_k^v}.
\label{transmission time}
\end{equation}

It is worth noting that the bandwidth allocation of service requests is highly relevant to the distances between the service requests and edge nodes. 
For short distances represented by a small $d_k^v$, a large bandwidth $b_k^v$ is generally not required. However, increasing $b_k^v$ becomes important to decrease the maximum transmission time when $d_k^v$ is significant.

\section{Problem Formulation}\label{section::problem}

In this section, we formulate the request scheduling problem as an Integer Linear Programming problem and prove its NP-hardness.
All symbols and variables are listed in Table~\ref{tab:var}.
\begin{table}[!tbp]
	\centering
	\normalsize
	\setlength\belowcaptionskip{0ex}
	\caption{Symbols and Variables}
	\label{tab:var}
	\renewcommand\arraystretch{1}   
	\begin{tabular*}{250pt}{ll}
		\toprule	
		\textbf{Symbols} & \textbf{Description}\\
		\midrule
		$\mathcal{G} = (\mathcal{V}, \mathcal{E})$ & Physical network graph\\
            $\mathcal{V}$ & Set of edge nodes\\
            $\mathcal{E}$ & Set of links\\
            $\mathcal{K}$ & Set of container types\\
            $\mathcal{T}$ & Set of time intervals\\
            $c_{k}$ & The required hardware resource \\
            & of type $k$ container\\
		$C^v$ & The hardware capacity of node $v$\\
		$B_v$ &The total bandwidth for edge node $v$ \\
        $L_k$ &The size of job for request $k$ \\
        $d^{v}_k$ & The distance between request $k$ \\
        & and node $v$\\
		\midrule
		
		\textbf{Variables} &\\
		$x_{k,t}^{v}$ & Binary variable whether request $k$\\
         &is assigned to node $v$ in time slot $t$\\
        $b^{v}_k$ & The bandwidth allocated to request $k$ \\
        & at node $v$\\

		\bottomrule
	\end{tabular*}
\end{table}

We consider two categories of latency that are important to the system performance, e.g., transmission latency and processing latency.

According to \cite{Ren2018}, \cite{Williams1991},
We formulate the average transmission latency as follows.

\begin{equation}
D_{tran} = \frac{L_k}{TS_v^k} x_{k,t}^v
\label{translatency}  
\end{equation}

where $L_k$ denotes the size of a job $k$.
We use $TS_v^k$ to represent the transmission rate. 

When a container processes a user request, a considerable amount of time is needed which is proportional to the size of the job.
The processing time of a user request in time slot $t$ is given by:
\begin{equation}
D_{proc} = \frac{L_k}{p_{k,t}} x_{k,t}^v
\label{proclatency}  
\end{equation}

where $p_{k,t}$ denotes the allocated processing capacity.
Let $x_{k,t}^v$ represent whether the request $k$ is distributed to base station $v$.


The overall latency of a user request can be formulated as:
\begin{equation}
D_{total} = D_{tran} + D_{proc} 
\label{coldlatency}  
\end{equation}

\textbf{Problem}
We provide the mathematical model of the container placement and allocation problem with bandwidth and processing capacity constraints, aiming to optimize the total latency of requests.

\begin{equation}
     \min max(\sum_{k \in \mathcal{K}} \sum_{v \in \mathcal{V}} \sum_{t \in T} D_{total} )
\label{optimization_equation_2}  \\  
\end{equation}
\begin{align}
    \text { s.t. } & 
     \sum_{k \in \mathcal{K}} b_{k}^v x_{k,t}^v  \leq B_{v}, \forall v \in \mathcal{V}, \forall t \in T  \label{bandwidthcon} \\
    &
     \sum_{k \in \mathcal{K}} c_{k} x_{k,t}^v  \leq C_{v},  \forall v  \in \mathcal{V}, \forall t \in T 
     \label{processingcon} \\
     &  
     \sum_{k \in \mathcal{K}} \sum_{v \in \mathcal{V}}  x_{k,t}^v = 1,
     \forall t \in T
     \label{integercon}
\end{align}

where $b_{k,t}$ denotes the allocated bandwidth of request $k$.
$B_{v}$ represents the total amount of bandwidth of base station $v$.
Also, we use $c_{k,t}$ to represent the amount of allocated processing capacity to run a container for request $n$.
Let $C_{v}$ denote the total amount of processing capacity at base station $v$.

Constraint~\ref{bandwidthcon} guarantees that the allocated bandwidth 
 of these containers must not exceed the total bandwidth capacity.
 Constraint~\ref{processingcon} guarantees that the allocated processing capacity must not exceed the total processing capacity.
Constraint~\ref{integercon} ensures that one request is only allocated once.

\subsection{Proof of NP-Hardness}
We show that the Generalized Assignment Problem (GAP), which is known to be NP-hard, can be reduced to the proposed problem.
The GAP problem is distributing a number of $K$ jobs to a set of $J$ agents with minimized cost.
For our problem, let $c_n$ be the size of the task.
Let each edge node $v$ represent an agent $j$ equipped with a resource capacity of $C_v$. Then, assigning a job to an agent becomes assign a serverless request to an edge node, aiming to minimize the end-to-end latency which can be mapped to the cost of GAP problem.
Thus, the GAP problem is a special case of our problem and hence our problem is NP-hard.
\vspace{-6pt}

\section{Request Scheduling}
\subsection{MIDACO}
MIDACO, a solver utilizing the Ant Colony algorithm, is suitable for commercial use in various numerical optimizations. It effectively handles continuous non-linear (NLP), discrete/integer (IP), and mixed integer (MINLP) optimization challenges~\cite{midaco}.

To solve the proposed problem, we formulate a set of internal constraints that any feasible
solution must also obey. These internal
constraints are shown in Equation~\ref{eq:mid-constraints}.

\begin{equation}
    \begin{aligned}
    & G(0) = sum(b_0^v) - B_v\\
    & G(1) = sum(b_1^v) - B_v\\ 
    &...\\
    & G(i-1) = sum(b_{i-1}^v) - B_v\\
    & G(i) = C_v - sum(c_i^v)\\
    & G(i+1) = C_v - sum(c_{i+1}^v)\\
    & ...\\
    & G(2i-1) = C_v - sum(c_{2i-1}^v),\\
    \end{aligned}
    \label{eq:mid-constraints}
\end{equation}
where G(0) to G(i-1) equals 0, and other G()$\ge$0. 

\subsection{Genetic Algorithm}
For this research, a custom Genetic Algorithm (GA) was designed to benchmark against other heuristic optimization methods. This GA combines bandwidth \{$b_k^v$\} with binary variables \{$x_k,t^v$\} to construct chromosomes. Initially, the GA generates a set of random parent chromosomes and pairs each to mate—randomly choosing a 'father' and 'mother' for crossover. The crossover occurs at a randomly chosen splice point, where father's genes up to the splice point are combined with the mother's genes from that point onward. For instance, a progeny chromosome at splice point $s$ inherits the father's genes $b_1^v$ through $b_s^v$ and the mother's genes $b_{s+1}^v$ through $b_n^v$, where $n$ is the chromosome's length. Post-crossover, the collective bandwidth may deviate from the limit. This is rectified by adjusting the genes so that the child's total bandwidth aligns with the preset constraint. Specifically, for the discrepancy $\Delta = sum\{child\_b_i\} - total \ bandwidth$, we alter a randomly chosen gene in the child's chromosome by incrementing or decrementing 1 depending on whether $\Delta$ is negative or positive, respectively. This adjustment is repeated for each unit of discrepancy, minus one.

After crossing, we perform mutation to introduce chromosome diversity.
We choose a random number $\mu \in (0,1)$ and use it to mutate the child items.
For bandwidth,
if $0.3 < \mu <  0.75$, we do not mutate.
If $\mu \ge  0.75$, we left-rotate the items one position.
If $\mu \le  0.5$, subtract 1 from the maximum item and add 1 to the minimum item 
in $child\_b_{i}$.

\subsection{CANN}
Both MIDACO and Genetic are heuristic algorithms, so a common issue for them is the converge time, i.e., how long do they cost to produce a stable solution. Normally, this process is counted by seconds, even minutes. Obviously, it is not practical to deploy such heuristic algorithms for the real system due to the fast dynamics. To solve this issue, we introduce a two-layer LSTM model to accelerate the process, which is trained by the results of MIDACO. With adequate training, the CANN is able to orchestrate with satisfied total delay and fast response time. To achieve the best performance, we always use the best result of MIDACO as training data, i.e., MIDACO 50000 in the experiment section.
\section{Experimental Evaluation}

\subsection{Experiment Setup}
The experiments were conducted on a system equipped with an AMD Ryzen 7 5800H processor, clocked at 3.20 GHz, and featuring 16GB RAM and 4 cores. We utilized MIDACO Version 6.0, licensed commercially, for mixed-integer programming and developed a custom genetic algorithm tailored for the research presented in this paper.

To assess the solvers' effectiveness, 100 scenarios with randomized service request locations were generated. In 
each scenario, there are 20 requests. Each request's hardware capacity needs are randomly assigned from a uniform range of 50 MHz to 150 MHz, and their proximity to the edge node follows a uniform distribution from 30 m to 200 m. There are 2 edge nodes, and each of them has 100 MHz bandwidth for data transmission and 1.5 GHz hardware capacity for data processing.

We evaluate 5 settings in the experiments. Genetic 5000 and 50000 means there are 5000 and 50000 generations in the Genetic algorithm respectively. MIDACO 5000 and 50000 means there are 5000 and 50000 times of evaluations in the algorithms respectively. For CANN, we use the output of MIDACO 50000 as the training lables.


\subsection{Simulation results}
MIDACO \cite{MIDACOSO37:online} is a solver for numerical optimization problems by using evolutionary hybrid algorithms.
Extensive numerical tests proved that MIDACO can achieve the global optimal solution on the majority of problems robustly\cite{MartinSchlueter2012}.
Hence, we adopt MIDACO to approach the near-optimal solutions of the proposed problem.

\begin{figure}[t]    \includegraphics[width=0.45\textwidth]{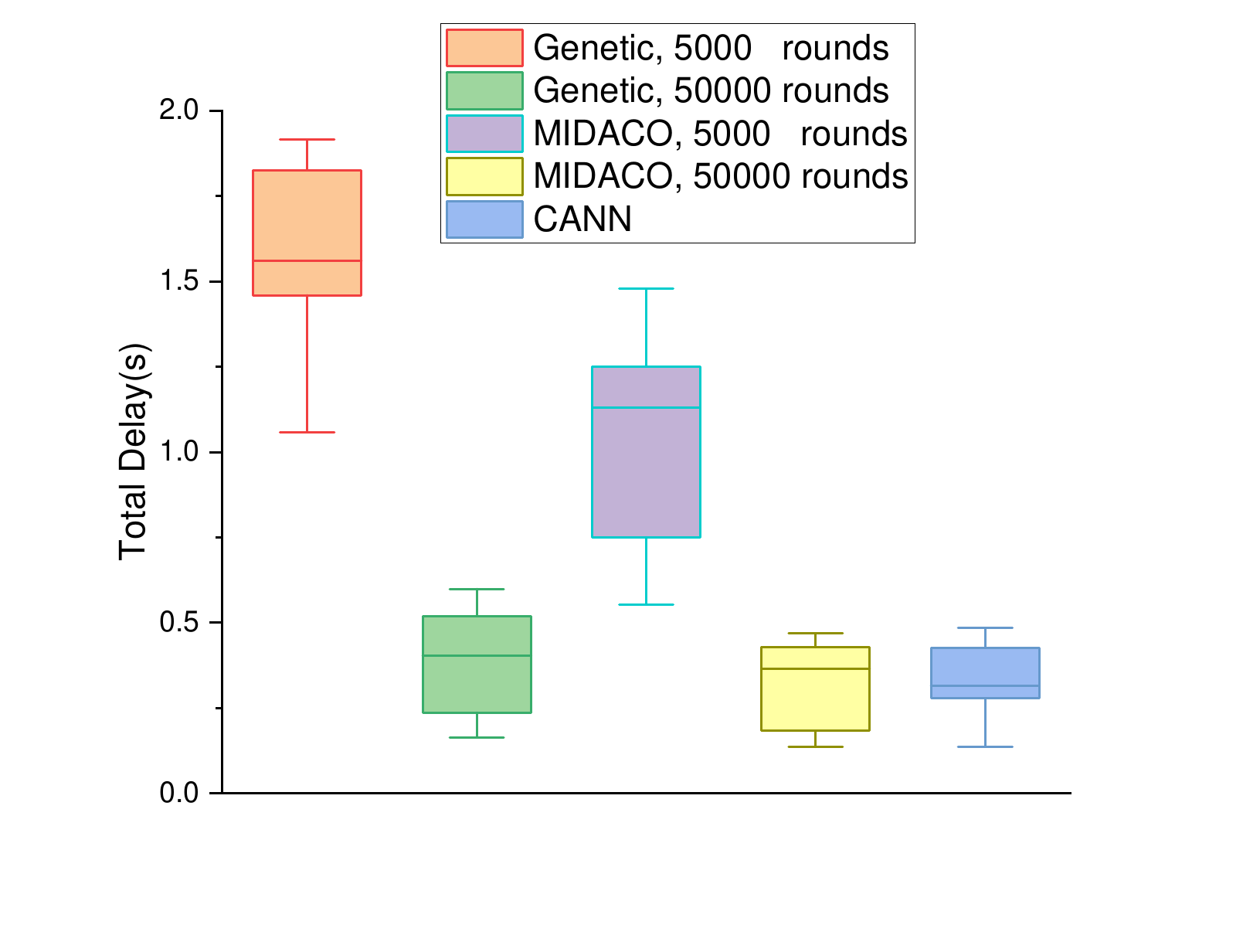}
    \centering
    \caption{Total delay of various settings}
    \label{total_delay}
\end{figure}

In Fig. \ref{total_delay}, the Genetic algorithm demonstrates a pronounced variability in delay outcomes contingent on the number of rounds. When operating with 5000 rounds, the Genetic algorithm incurs a substantial delay of approximately 1.5 seconds. However, when the number of rounds is increased to 50000, there is a noteworthy decrease in total delay, reducing to about 0.42 seconds. This reduction of 1.08 second signifies an enhanced performance attributed to a more exhaustive search capability, allowing the algorithm to explore and refine solutions more effectively over a larger number of generations.

Conversely, the MIDACO solver exhibits a lesser degree of improvement in delay reduction when comparing its two scales of operation. At 5000 rounds, the MIDACO solver records a delay of around 1.17 second, which marginally decreases to just below 0.5 second at 50000 rounds. This minimal decrease suggests that while MIDACO benefits from increased rounds, the scale of improvement is significantly tapered, indicating a potential saturation in efficiency gains beyond a certain number of rounds.

\begin{figure}[t]
    \includegraphics[width=0.45\textwidth]{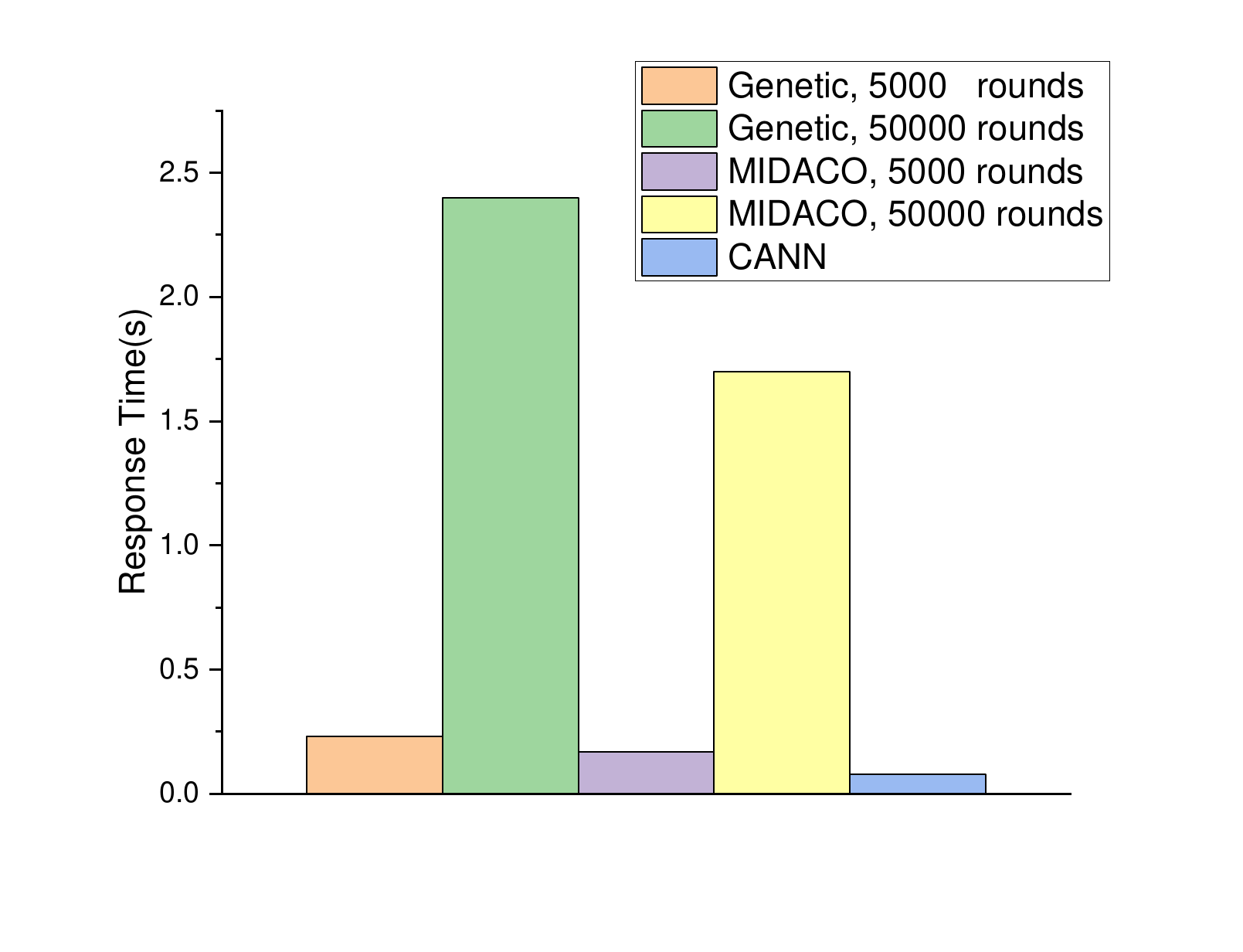}
    \centering
    \caption{Response time of various settings}
    \label{response time}
\end{figure}

Since both MIDACO and genetic are heuristic, it is essential to compare their response time to CANN. 
Fig. \ref{response time} provides a detailed comparative analysis of convergence times associated with the Genetic algorithm, MIDACO solver, and CANN across different scales of computational effort.

Analyzing the data, the Genetic algorithm at 5000 generations manifests a convergence time of approximately 2.0 seconds, indicating moderate efficiency under constrained iterations. However, when the generations are increased to 50000, the convergence time reduces significantly to about 0.5 seconds, a fourfold decrease. This substantial improvement highlights the Genetic algorithm's capability to optimize solutions more effectively with greater computational leeway, suggesting its suitability for complex problems where more extensive solution exploration is feasible.

In contrast, the MIDACO solver displays a convergence time of around 1.5 seconds at 5000 evaluations. When the evaluations are escalated to 50000, the time slightly decreases to 1.0 seconds. 
The CANN model shows a convergence time slightly below 1.0 second, outperforming the other algorithms at their lower operational settings and closely matching the MIDACO at its higher setting.

\begin{figure}[t]
    \includegraphics[width=0.45\textwidth]{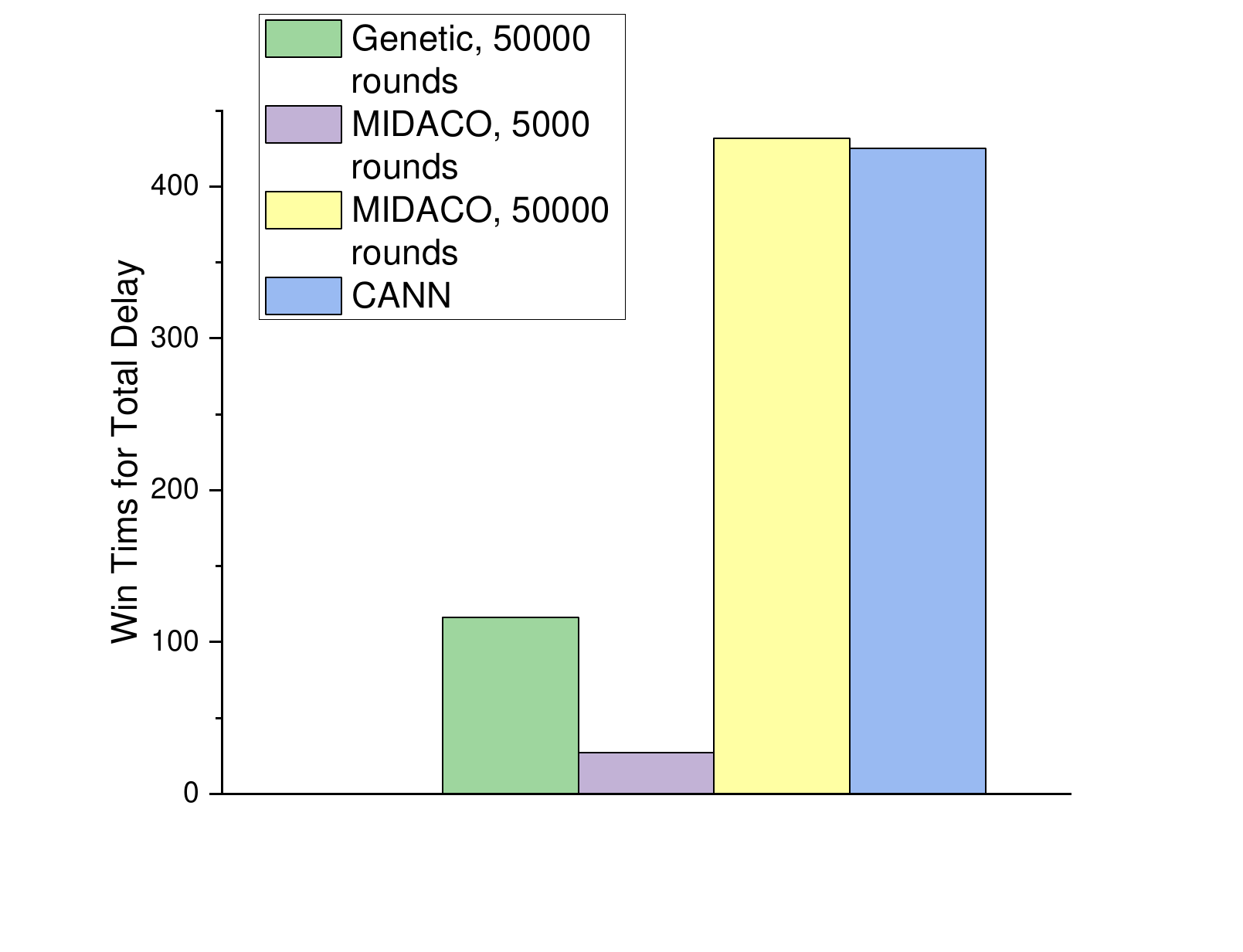}
    \centering
    \caption{How many times the various settings win for total delay}
    \label{win times}
\end{figure}

Fig. \ref{win times} indicates the details about the championship of various settings for the total delay based on 1000 samples. It is notable that Genetic 5000 never gets a championship. For the MIDACO, it is possible to be the winner in some circumstances. The number of championships for MIDACO 50000 and CANN is almost the same, around 420. 
\section{Conclusion}\label{conclusion}
In this paper, we have investigated the container placement problem in serverless edge computing.
By jointly considering wireless bandwidth and node capacity allocation, we have proposed a context-aware NN to reduce the end-to-end latency. 
Experimental results have shown that the CANN achieves comparable performance on the total delay and times of winning, whereas the converge speed is much better than MIDACO and Genetic, especially when the iteration of heuristics increases dramatically.


\section*{Acknowledgment}
This work
was supported by the Norwegian Research Council under
Grant 262854/F20 (DILUTE project).

\bibliographystyle{unsrt}
{
\bibliography{main}}

\end{document}